\documentclass[pra,aps,superscriptaddress,twocolumn,color]{revtex4-1}

\usepackage{amsmath}
\usepackage{amsfonts}
\usepackage{bm}
\usepackage{graphicx}
\usepackage{array,color}

\begin{document}
\title{Symmetry-breaking instability of leapfrogging vortex rings in a
Bose-Einstein condensate}

\author{Mayumi Ikuta}
\affiliation{Department of Engineering Science, University of Electro-Communications, Tokyo 182-8585, Japan}

\author{Yumi Sugano}
\affiliation{Department of Engineering Science, University of Electro-Communications, Tokyo 182-8585, Japan}

\author{Hiroki Saito}
\affiliation {Department of Engineering Science, University of Electro-Communications, Tokyo 182-8585, Japan}

\date{\today}
\begin{abstract}
Three coaxial quantized vortex rings in a Bose-Einstein condensate exhibit
aperiodic leapfrogging dynamics.
It is found that such circular vortex rings are dynamically unstable
against deformation breaking axial rotational symmetry.
The dynamics of the system are analyzed using the Gross-Pitaevskii and
vortex-filament models.
The dependence of the instability on the initial arrangement of the vortex
rings is investigated.
The system is found to be significantly unstable for a specific
configuration of the three vortex rings.
\end{abstract}

\maketitle

\section{Introduction}
\label{s:introduction}

A vortex ring has a toroidal vorticity distribution with a torus shape, and
survives for a long time after it is created.
A vortex ring has a linear momentum and can therefore travel a long
distance, as for the well-known example of smoke rings in air.
However, in a normal fluid with viscosity, such as air or water,
a vortex ring eventually decays due to the dissipation of energy and
momentum.
Furthermore, the toroidal vorticity distribution has an inherent instability
against azimuthal-wave excitation~\cite{Widnall,Fukumoto}.

In superfluids, the situation is much simpler.
Vortices are quantized, and a vortex ring is simply a circular closed
loop of a quantized vortex line.
Because of the absence of viscosity, a circular quantized vortex ring
has an infinite lifetime in a uniform superfluid.
Quantized vortex rings were first detected in superfluid helium using ion
spectroscopy~\cite{Rayfield}.
In Bose-Einstein condensates (BECs) of ultracold atomic gases, quantized
vortex rings have been created through the decay of solitons and directly
observed by imaging the density
distribution~\cite{Anderson,Ginsberg,Shomroni,Becker}.
The dynamics of quantized vortex rings in superfluids have been
theoretically studied by many researchers~\cite{Koplik, Jackson, Komineas,
  Crasovan, Komineas2, Horng, Helm, Reichl, Wacks, Wacks2, Caplan, Bisset,
  Gubser, Zhu, Wang, Ruban, Ticknor}.

A circular vortex ring in a superfluid travels at a constant velocity
roughly proportional to the inverse of its radius.
Such a quantized vortex ring traveling in a uniform superfluid is stationary
in the moving frame of reference, and stable against perturbations.
When two circular vortex rings with the same vorticity are arranged in
parallel sharing the same axis, the radius of the front (rear) ring
increases (decreases) and the ring decelerates (accelerates).
The rear ring thus passes through the front ring, and then the roles of the
two rings are reversed, which results in leapfrogging dynamics of the two
vortex rings~\cite{Helmholtz,Mele}.
For more than two coaxial vortex rings, the leapfrogging dynamics become
more complicated.

In the present paper, we focus on the axial-symmetry breaking instability of
leapfrogging vortex rings in superfluids.
When circular vortex rings are coaxially arranged in the initial state, the
system has rotational symmetry about the axis of the rings.
In the ensuing leapfrogging dynamics, the axial symmetry is spontaneously
broken, i.e., infinitesimal azimuthal perturbations are exponentially
increased, and the leapfrogging dynamics are eventually destroyed by
significant distortions of the rings.
Such instabilities of coaxial vortex rings have been studied using the
vortex-filament model.
In Ref.~\cite{Wacks}, the dynamics of leapfrogging vortex rings were
numerically studied and axial-symmetry breaking instability was shown to
occur for seven vortex rings.
A linear stability analysis was performed for two vortex rings by
Ref.~\cite{Gubser}.
The long-time stability of two and three vortex rings was studied by
Ref.~\cite{Ruban}.

In the present paper, we investigate the axial-symmetry breaking instability
of three quantized vortex rings using both the Gross-Pitaevskii (GP)
equation and the vortex-filament model.
Three vortex rings with the same direction of vorticity are coaxially
arranged, which exhibit leapfrogging dynamics.
We find that the system is significantly unstable against axial-symmetry
breaking for a specific initial arrangement of the vortex rings.
The instability occurs both for the GP and vortex-filament models.
We investigate the dependence of the instability on the initial
arrangements of the vortex rings and on the interaction coefficient.
We will show that the symmetry-breaking modulation significantly grows
when the vortex rings form a particular configuration during the dynamics,
and most unstable eigenmode is obtained.

This paper is organized as follows.
Section~\ref{s:GP} presents a study of the dynamics of vortex rings by
solving the GP equation numerically.
Section~\ref{s:filament} analyzes the symmetry-breaking instability using
the vortex-filament model.
Section~\ref{s:conc} gives the conclusions of this study.

\section{Gross-Pitaevskii model}
\label{s:GP}

We consider a BEC of a dilute atomic gas at zero temperature described by
the GP equation,
\begin{equation}
i \hbar \frac{\partial \psi}{\partial t} = -\frac{\hbar^2}{2M}
\nabla^2 \psi + V(\bm{r}) \psi + \frac{4\pi\hbar^2 a}{M} |\psi|^2 \psi,
\end{equation}
where $\psi(\bm{r}, t)$ is the macroscopic wave function of the condensate,
$M$ is the mass of an atom, $V(\bm{r})$ is an external potential, and $a$ is
the $s$-wave scattering length.
We normalize the length, time, energy, and atomic density by an arbitrary
length $L$, arbitrary time $T$, $\hbar / T$, and arbitrary density $n_0$,
respectively, where $\hbar T = M L^2$ is satisfied.
The GP equation then becomes non-dimensional as
\begin{equation} \label{GP}
i \frac{\partial \psi}{\partial t} = -\frac{\nabla^2}{2}
\psi + V(\bm{r}) \psi + g |\psi|^2 \psi,
\end{equation}
where the non-dimensional interaction coefficient is $g = 4 \pi a L^2 n_0$.
To reduce the effect of boundary conditions on the axial symmetry of the
vortex rings, we use a cylindrical tube potential given by
\begin{equation} \label{V}
  V(\bm{r}) = \left\{ \begin{array}{ll} 0 & (r_\perp < R_{\rm wall}) \\
    V_0 & (r_\perp \geq R_{\rm wall}) \end{array} \right.,
\end{equation}
where $r_\perp = (x^2 + y^2)^{1/2}$ and $R_{\rm wall}$ is the radius of the
cylindrical tube.
Such a potential can be produced by a phase-imprinted laser
beam~\cite{Gaunt}.
The height $V_0$ of the potential wall is taken to be much larger than the
chemical potential $g|\psi|^2$.
A periodic boundary condition is imposed in the $z$ direction.

The initial state is prepared by the imaginary-time propagation of
Eq.~(\ref{GP}), where $i$ on the left-hand side is replaced with $-1$.
The wave function is normalized with the volume of the cylindrical tube as
\begin{equation}
  \int |\psi|^2 d\bm{r} = \pi R_{\rm wall}^2 L_z,
\end{equation}
where $L_z$ is the size of the system in the $z$ direction.
The density $|\psi|^2$ is thus almost unity inside the tube potential, when
the radius $R_{\rm wall}$ is much larger than the healing length.
After the imaginary-time propagation converges, circular vortex rings are
imprinted in such a way that their symmetry axis is on the $z$ axis.
The wave function is multiplied by $\exp[i \phi(\bm{r})]$, with the phase
given by
\begin{eqnarray} \label{phi}
\phi(\bm{r}) & = & \sum_{j=1}^{N_{\rm ring}} \sum_{n = -n_z}^{n_z} \biggl(
\tan^{-1} \frac{z - Z_j - n L_z}{r_\perp + R_j}
\nonumber \\
& & - \tan^{-1} \frac{z - Z_j - n L_z}{r_\perp - R_j} \biggr),
\end{eqnarray}
where $N_{\rm ring}$ is the number of vortex rings and $R_j$ and $Z_j$
are the radius and $z$ coordinate of the $j$th vortex ring, respectively.
The first and second terms in the bracket in Eq.~(\ref{phi}) cancel the
radial flow $\partial \phi / \partial r_\perp$ on the $z$ axis, and the
summation over integers $n$ in Eq.~(\ref{phi}) ensures a periodic 
boundary condition in the $z$ direction for a sufficiently large $n_z$,
which reduces the initial disturbances due to the phase imprint.
To further reduce the initial disturbances, we perform a short
imaginary-time evolution during $T_{\rm imag}$ after the phase
$\phi(\bm{r})$ is imprinted on the wave function.
This imaginary-time evolution slightly changes the radii and $z$ positions
of the vortex rings from $R_j$ and $Z_j$ in Eq.~(\ref{phi}).

We obtain the imaginary and real time evolution of the system by numerically
solving Eq.~(\ref{GP}) using the pseudospectral method~\cite{recipe}.
The numerical mesh is typically $512^3$ with a spatial discretization of $dx
= dy = dz = 0.25$, and the system size is $L_x \times L_y \times L_z = 128^3$.
Small noise must be added to the initial state to trigger the symmetry
breaking instability.
To generate moderate noise, we first set complex random numbers with a
normal distribution on each numerical mesh point and eliminate large
wavenumber components with $k > k_{\rm cutoff}$ using a Fourier transform,
where $k_{\rm cutoff}$ is taken to be $10 \times 2 \pi / 128$.
Such low-pass filtered noise multiplied by a small number (typically 0.1) is
added to the wave function after the imaginary-time evolution for
$T_{\rm imag}$.
In the following calculations, we take $V_0 = 10$, $R_{\rm wall} = 63$,
$n_z = 50$, and $T_{\rm imag} = 4$.

\begin{figure}[tb]
\includegraphics[width=8cm]{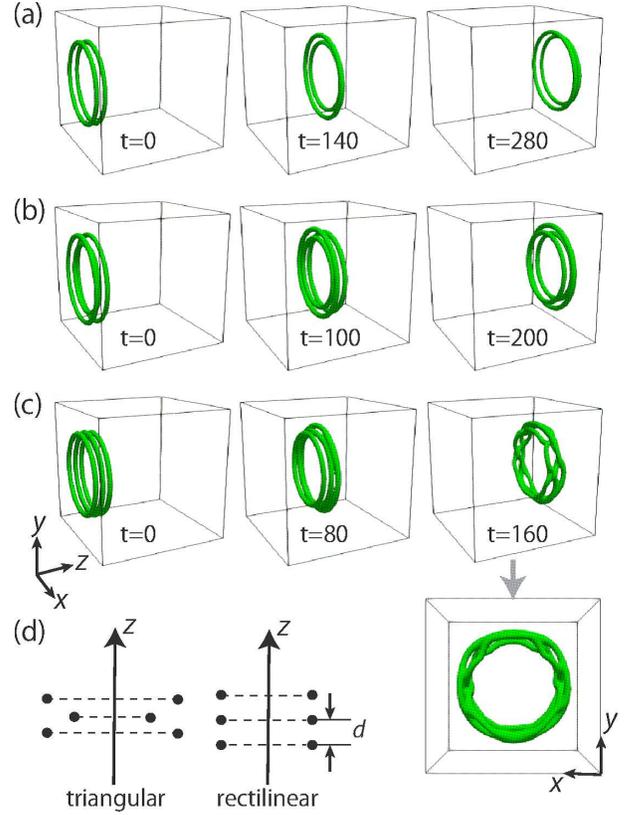}
\caption{
  (color online) (a)-(c) Dynamics of vortex rings obtained by solving the GP
  equation with $g = 1$.
  The isodensity surface of the density $|\psi|^2 = 0.5$ is shown.
  (a) Two vortex rings initially located with $R_1 = R_2 = 20$, $Z_1 = -28$,
  and $Z_2 = -26$.
  (b) Three vortex rings with a triangular initial arrangement, where $R_1 =
  20 - 2 \sqrt{3}/3$, $R_2 = R_3 = 20 + 1/\sqrt{3}$, $Z_1 = -26$, $Z_2 =
  -28$, and $Z_3 = -24$.
  (c) Three vortex rings with a rectilinear initial arrangement, where $R_1
  = R_2 = R_3 = 20$, $Z_1 = -27$, $Z_2 = -26$, and $Z_3 = -28$.
  The vortices seen from the $+z$ direction at $t = 160$ are highlighted.
  Axial symmetry is retained in (a) and (b) and broken in (c) in the time
  evolution.
  The size of the box is $64^3$, with the origin at the center.
  See the Supplemental Material for movies showing the
  dynamics~\cite{movies}.
  (d) Initial vortex arrangements set by Eq.~(\ref{phi}).
  The triangular and rectilinear arrangements used in (b) and (c)
  are shown in the left-hand and right-hand panels, respectively, where the
  dots represent the positions of vortex cores on the plane including the $z$
  axis.
}
\label{f:gpevo}
\end{figure}
Figure~\ref{f:gpevo} shows the time evolution of two and three vortex rings
(see the Supplemental Material for movies~\cite{movies}).
Figure~\ref{f:gpevo}(a) shows the case of two vortex rings that are
coaxially arranged at $t = 0$.
They exhibit periodic leapfrogging dynamics and travel in the $+z$
direction, in which the axial symmetry is retained.
Figure~\ref{f:gpevo}(b) shows the case of three vortex rings coaxially
arranged so that the six vortex cores form two regular triangles on
the plane including the symmetry axis, as illustrated in the left-hand
panel in Fig.~\ref{f:gpevo}(d).
We refer to such an initial arrangement of vortex rings as a ``triangular''
arrangement.
The rings in this arrangement also exhibit periodic leapfrogging dynamics
such that the triangles of the vortex cores rotate around their centers.
The axial symmetry is retained for a long time in the leapfrogging dynamics
for the triangular configuration, as reported for the vortex-filament
model~\cite{Wacks,Ruban}.
We have numerically confirmed that the axially-symmetric leapfrogging
dynamics continue at least until $t \simeq 500$ for Figs.~\ref{f:gpevo}(a)
and \ref{f:gpevo}(b).

\begin{figure}[tb]
\includegraphics[width=8cm]{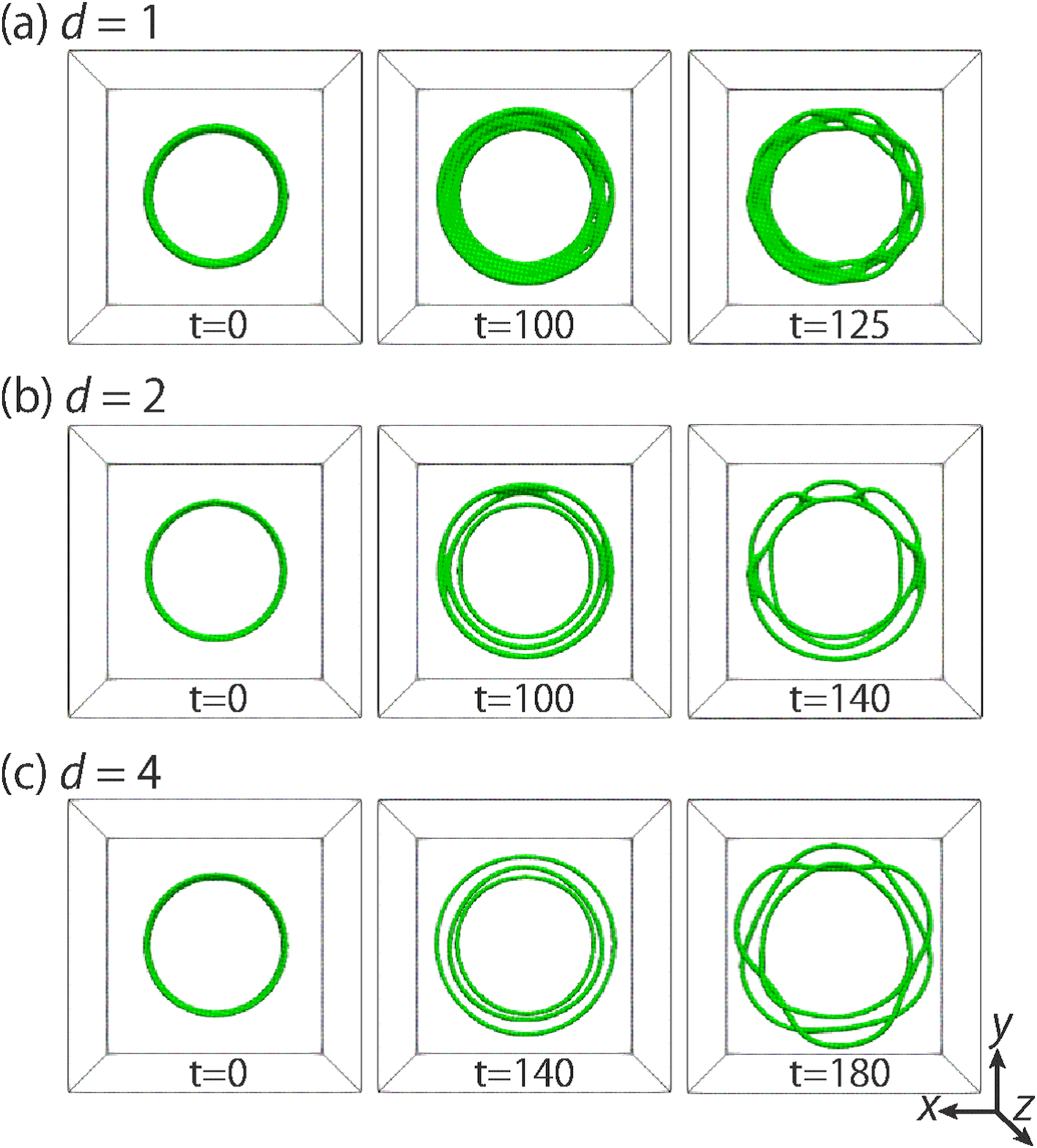}
\caption{
  (color online) Dependence of the symmetry breaking dynamics on the initial
  distance $d$ between the vortex rings in the rectilinear initial
  arrangement with radius $R = 20$.
  The dynamics are obtained by solving the GP equation with $g = 1$.
  (a) $d = 1$, (b) $d = 2$, and (c) $d = 4$.
  The dynamics in (b) is the same as that in Fig.~\ref{f:gpevo}(c).
  The isodensity surface of the density $|\psi|^2 = 0.2$ is shown.
  The size of the box is $64^3$, with the origin at the center, and is
  seen from the $+z$ direction.
  See the Supplemental Material for movies showing the
  dynamics~\cite{movies}.
}
\label{f:d-dep}
\end{figure}
In Fig.~\ref{f:gpevo}(c), three vortex rings with the same radius are
coaxially and equidistantly arranged along the $z$ axis.
We refer to such an initial arrangement as a ``rectilinear'' arrangement.
The leapfrogging dynamics is irregular for this initial state.
We label the vortex rings A, B, and C from front to rear in the initial
arrangement.
First, B and C pass through A, and the order is reversed as C, B, A from
front to rear.
Then A and B pass through C, during which a leapfrog occurs between A and B,
resulting in the order B, A, C from front to rear.
Next, A is overtaken by C, which gives the snapshot at $t = 80$ in
Fig.~\ref{f:gpevo}(c).
We can see that the axial symmetry starts to break at this time.
Subsequently, the axial symmetry breaking develops as irregular leapfrogging
dynamics.
The vortex lines then intertwine with each other and reconnections occur.

Figure~\ref{f:d-dep} shows the dynamics seen from the $+z$ direction for the
rectilinear initial arrangement with various distances $d$ between the
vortex rings.
We see that the axial symmetry is broken at $t \sim 100$, and the unstable
modes depend on the initial distance $d$.
The wavelengths of the unstable modes are $\simeq 2 \pi R / 8$ in
Fig.~\ref{f:d-dep}(a), $\simeq 2 \pi R / 4$ in Fig.~\ref{f:d-dep}(b), and
$\simeq 2 \pi R / 3$ in Fig.~\ref{f:d-dep}(c) with $R$ being the radius of
the vortex ring, which indicates that the most unstable wavelength increases
with $d$.
The growth rate for the unstable mode tends to decrease with increasing
$d$.
The distances indicated in Fig.~\ref{f:d-dep} are the distances between the
vortex rings in the phase imprint by Eq.~(\ref{phi}).
By the imaginary-time evolution for $T_{\rm imag} = 4$ after the phase
imprint, the distance between adjacent vortices changes from $d = 1$, 2,
and 4 to $d \simeq 2$, 3.5, and 5, respectively.

\begin{figure}[tb]
\includegraphics[width=8cm]{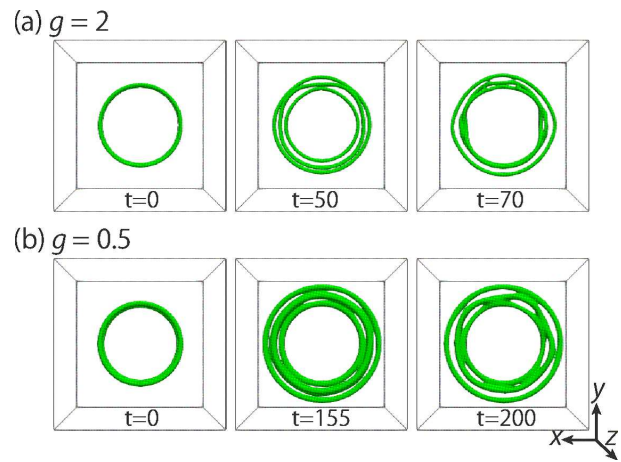}
\caption{
  (color online) Dependence of symmetry-breaking dynamics on
  interaction coefficient $g$ for vortex rings in rectilinear
  initial arrangement with $d = 2$ and $R = 20$.
  The dynamics are obtained by solving the GP equation.
  (a) $g = 2$ and (b) $g = 0.5$.
  The isodensity surfaces of the densities $|\psi|^2 = 0.5$ and 0.2 are
  shown for (a) and (b), respectively.
  The size of the box is $64^3$ with the origin at the center, and is
  seen from the $+z$ direction.
  See the Supplemental Material for movies showing the
  dynamics~\cite{movies}.
}
\label{f:g-dep}
\end{figure}
Figure~\ref{f:g-dep} shows the dependence of the symmetry-breaking dynamics
on the interaction coefficient $g$.
For $g = 2$, the wavelength of the unstable mode is $\simeq 2 \pi R / 4$ (or
$\simeq 2 \pi R / 3$), as shown in Fig.~\ref{f:g-dep}(a).
For $g = 0.5$, the unstable wavelength is $\simeq 2 \pi R / 5$, as shown in
Fig.~\ref{f:g-dep}(b).
This difference is purely due to the difference in $g$, because the distance
$d$ between the vortices after the imaginary-time propagation is almost the
same for Figs.~\ref{f:g-dep}(a) and \ref{f:g-dep}(b).
For a larger value of $g$, the healing length becomes smaller, and
therefore, the spatial discretization $dx$, $dy$, and $dz$ must be decreased
and the numerical mesh must be increased.
For a smaller value of $g$, the time for which the instability emerges
increases and long-time evolution is needed.

The symmetry-breaking instability demonstrated in
Figs.~\ref{f:gpevo}-\ref{f:g-dep} is a modulational instability (or
dynamical instability), in which infinitesimal symmetry-breaking noise grows
exponentially with time.
However, in the present case, the standard linear stability analysis, i.e.,
Bogoliubov analysis, cannot be used, since the vortex rings move and the
system is not in a stationary state.
We also cannot use the Floquet analysis, since the dynamics of the three
vortex rings are not periodic for the rectilinear initial arrangement.

\section{Vortex-filament model}
\label{s:filament}

To study the symmetry-breaking dynamics of vortex rings in more detail, we
employ the vortex-filament model.
In this model, we focus only on the dynamics of quantized
vortex lines.
We assume that the fluid is incompressible and irrotational except at the
vortex lines.
A vortex segment located at $\bm{r}$ moves with the velocity given by the
Biot-Savart law,
\begin{equation} \label{BS}
\bm{v}(\bm{r}) = \frac{\kappa}{4\pi} \int \frac{(\bm{r}' - \bm{r}) \times
  d\bm{r}'}{|\bm{r} - \bm{r}'|^3},
\end{equation}
where $\kappa = 2\pi$ is the normalized circulation of a quantized vortex
and the line integral is taken along all vortex lines.

In the numerical calculations, the $n$th vortex ring is represented by a
sequence of $N_p$ positions on the ring, $\bm{S}_1^{(n)}$, $\bm{S}_2^{(n)}$,
$\cdots$, $\bm{S}_{N_p}^{(n)}$, and $\bm{S}_{N_p+1}^{(n)} = \bm{S}_1^{(n)}$.
The Biot-Savart integral in Eq.~(\ref{BS}) is then rewritten as
\begin{eqnarray}
  \bm{v}(\bm{r}) & = & \frac{\kappa}{4\pi} \sum_{n=1}^{N_{\rm ring}}
  \sum_{j=1}^{N_p} \int_{\bm{S}_j^{(n)}}^{\bm{S}_{j+1}^{(n)}}
  \frac{(\bm{r}' - \bm{r}) \times d\bm{r}'}{|\bm{r} - \bm{r}'|^3}
  \nonumber \\
  & \equiv & \frac{\kappa}{4\pi} \sum_{n=1}^{N_{\rm ring}}
  \sum_{j=1}^{N_p} F_j^{(n)}(\bm{r}).
  \label{BS2}
\end{eqnarray}
We approximate the vortex line between $\bm{S}_j^{(n)}$ and
$\bm{S}_{j+1}^{(n)}$ to be a straight line~\cite{Schwarz, Barenghi}, and the
line integral becomes
\begin{equation} \label{BS3}
F_j^{(n)}(\bm{r}) = \frac{(D_j^{(n)} + D_{j+1}^{(n)})
    (\bm{D}_j^{(n)} \times \bm{D}_{j+1}^{(n)})}{D_j^{(n)} D_{j+1}^{(n)}
  (D_j^{(n)} D_{j+1}^{(n)} + \bm{D}_j^{(n)} \cdot \bm{D}_{j+1}^{(n)})},
\end{equation}
where $\bm{D}_j^{(n)} = \bm{S}_j^{(n)} - \bm{r}$.
Since the vortex-filament model breaks down for $|\bm{r} - \bm{r}'|$ smaller
than the size of the vortex core $\xi$, we omit the line integral inside the
vortex core, $|\bm{r}' - \bm{r}| < \xi$, in Eqs.~(\ref{BS2}) and
(\ref{BS3}), which avoids the divergence at $\bm{r}' = \bm{r}$.
The length $\xi$ corresponds to the healing length in the GP model and we
take $\xi = 1$, corresponding to $g = 1$.
The number of points per vortex ring is taken to be $N_{\rm ring} = 256$.
Because the radii of the rings that we are considering are $\simeq 20$, the
distance between the adjacent points $|\bm{S}_j^{(n)} - \bm{S}_{j+1}^{(n)}|$
is typically $2\pi \times 20 / 256 \simeq 0.5 \lesssim \xi$.
This is in contrast to the case of liquid helium~\cite{Schwarz}, in which
the size of the vortex core is $\sim 10^{-10}$ m and usually much smaller
than the distance between the adjacent points $|\bm{S}_j^{(n)} -
\bm{S}_{j+1}^{(n)}|$.
We do not implement vortex reconnections in our calculations, because we
focus on the axial-symmetry breaking.

The dynamics of $N_{\rm ring} \times N_p$ points are obtained by numerically
solving the equation of motion $d{\bm{S}_j^{(n)}} / dt =
\bm{v}(\bm{S}_j^{(n)})$ using the fourth-order Runge-Kutta method.
The initial positions of the points are set to
\begin{equation}
  \bm{S}_j^{(n)} = R_n \bm{r}_j + Z_n \bm{z},
\end{equation}
where $R_n$ and $Z_n$ are the radius and $z$-coordinate of the $n$th vortex
ring, $\bm{r}_j = \bm{x} \cos(2\pi j / N_p) + \bm{y} \sin(2\pi j / N_p)$ is
the unit vector in the radial direction, and $\bm{x}$, $\bm{y}$, and
$\bm{z}$ are the unit vectors in Cartesian coordinate.
Small initial noise is added to each point to trigger the axial symmetry
breaking.
To avoid numerical instability in the vortex-filament model, we cut off
the large wavenumber components in each time step as follows.
We perform a Fourier transform of the radius and $z$-coordinate as
\begin{eqnarray}
 \label{cmn}
  c_m^{(n)} & = & \sum_{j=1}^{N_p} \sqrt{(\bm{S}_j^{(n)} \cdot \bm{x})^2 +
    (\bm{S}_j^{(n)} \cdot \bm{y})^2} e^{-i m j}, \\
 \label{dmn}
  d_m^{(n)} & = & \sum_{j=1}^{N_p} \bm{S}_j^{(n)} \cdot \bm{z} e^{-i m j},
\end{eqnarray}
where $m$ is an integer and $c_{-m}^{(n)} = c_m^{(n)*}$ and $d_{-m}^{(n)} =
d_m^{(n)*}$ are satisfied.
We then eliminate Fourier components with $|m|$ larger than $m_{\rm cutoff}$, 
and perform an inverse Fourier transform as
\begin{equation} \label{invF}
  \bm{S}_j^{(n)} = N_p^{-1} \sum_{|m| \leq m_{\rm cutoff}} \left(c_m^{(n)}
  \bm{r}_j + d_m^{(n)} \bm{z} \right) e^{i m j}.
\end{equation}
We take $m_{\rm cutoff} = 10$ in the following calculations.

\begin{figure}[tb]
\includegraphics[width=8cm]{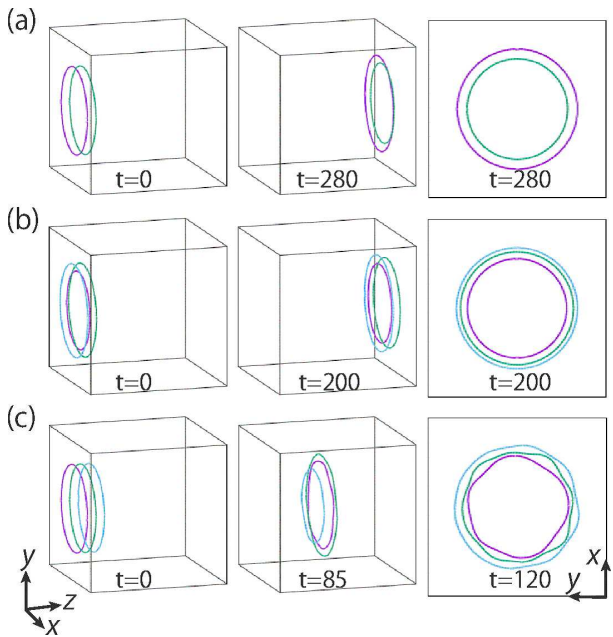}
\caption{
  (color online) Dynamics of vortex rings obtained by the vortex-filament
  model.
  (a) Two vortex rings initially located with $R_1 = R_2 = 20$, $Z_1 = -30$,
  and $Z_2 = -26$.
  (b) Three vortex rings with a triangular initial arrangement, where $R_1 =
  20 - 4 \sqrt{3}/3$, $R_2 = R_3 = 20 + 2/\sqrt{3}$, $Z_1 = -28$, $Z_2 =
  -26$, and $Z_3 = -30$.
  (c) Three vortex rings with a rectilinear initial arrangement, where $R_1
  = R_2 = R_3 = 20$, $Z_1 = -30$, $Z_2 = -26$, and $Z_3 = -22$.
  Vortices are seen from the $+z$ direction in the rightmost panels.
  Axial symmetry is retained in (a) and (b) and broken in (c).
  The size of the box is $64^3$ with the origin at the center.
  See the Supplemental Material for movies showing the
  dynamics~\cite{movies}.
}
\label{f:filament}
\end{figure}
Figure~\ref{f:filament} shows the dynamics of vortex rings.
In Fig.~\ref{f:filament}(a), two vortex rings with the same radius $R_1 =
R_2 = 20$ are coaxially arranged at a distance $Z_2 - Z_1 = 4$ in the initial
state, and they exhibit periodic leapfrogging dynamics in the time
evolution.
The frequency of the leapfrogging dynamics in Fig.~\ref{f:filament}(a) is
similar to that in the GP model in Fig.~\ref{f:gpevo}(a).
The period of the leapfrogging dynamics is roughly proportional to the
distance between the vortex rings.
In the GP model in Fig.~\ref{f:gpevo}(a), the distance is $Z_2 - Z_1 = 2$ in
the initial phase imprint, though this increases during the
imaginary-time evolution for $T_{\rm imag}$, which is why the leapfrog
frequency in Fig.~\ref{f:gpevo}(a) is similar to that in
Fig.~\ref{f:filament}(a) with the initial distance $Z_2 - Z_1 = 4$.
The vortex-core structure in the GP model may also affect the leapfrog
frequency, which is not included in the vortex-filament model.

Figures~\ref{f:filament}(b) and \ref{f:filament}(c) show the dynamics of
three vortex rings with triangular and rectilinear initial
arrangements, respectively.
For the triangular initial arrangement, the three vortex rings exhibit
almost periodic leapfrogging dynamics and the axial symmetry is retained
until $t = 200$, as shown in Fig.~\ref{f:filament}(b).
For the rectilinear initial arrangement, in contrast, the axial symmetry
is broken and modulation arises after a few leapfrogs, as shown in
Fig.~\ref{f:filament}(c).
This behavior of the vortex rings is similar to that in the GP model in
Fig.~\ref{f:gpevo}.

\begin{figure}[tb]
\includegraphics[width=8cm]{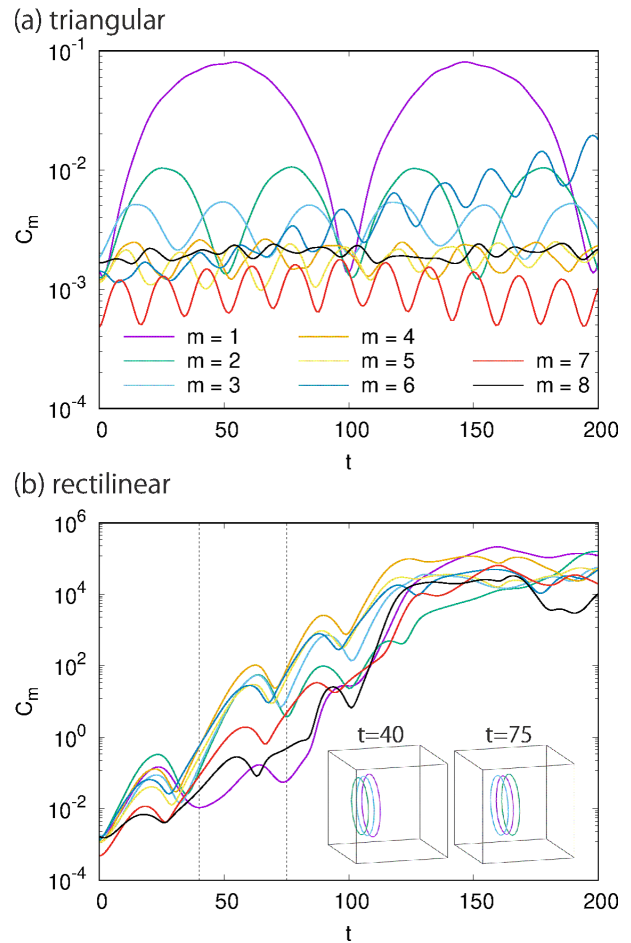}
\caption{
  (color online) Time evolution of the Fourier components $C_m$ defined in
  Eq.~(\ref{cm}).
  (a) and (b) correspond to the dynamics in Figs.~\ref{f:filament}(b) and
  \ref{f:filament}(c), respectively.
  The vertical dashed lines in (b) indicate $t = 40$ and $t = 75$, for which
  $C_m$ is exponentially rising.
  The insets in (b) show the vortex rings at these times.
}
\label{f:fourier}
\end{figure}
The degree of axial symmetry breaking is quantified by the Fourier
coefficients $c_m^{(n)}$ and $d_m^{(n)}$ in Eqs.~(\ref{cmn}) and (\ref{dmn})
for $m \neq 0$.
We define the degree of axial symmetry breaking for mode $m$ as
\begin{equation} \label{cm}
  C_m = \sum_{n=1}^{N_{\rm ring}} \left(|c_m^{(n)}|^2 + |d_m^{(n)}|^2
  \right).
\end{equation}
Figures~\ref{f:fourier}(a) and \ref{f:fourier}(b) show the time evolution of
$C_m$ for the dynamics in Figs.~\ref{f:filament}(b) and \ref{f:filament}(c),
respectively.
The initial values of $C_m \sim 10^{-3}$ originate from the initial random
noise.
For the triangular initial arrangement, the values of $C_m$ are suppressed
below $10^{-1}$ until $t = 200$, as shown in Fig.~\ref{f:fourier}(a).
They oscillate and never grow for $m \neq 6$.
The growth of the $m = 6$ mode is slow and only affects the long-time
dynamics~\cite{Ruban}.
The time evolution of $C_m$ for the rectilinear initial
arrangement in Fig.~\ref{f:fourier}(b) is qualitatively different from
that in Fig.~\ref{f:fourier}(a).
The values of $C_m$ exponentially grow in time as they oscillate,
reflecting the fact that axial symmetry breaking occurs in
Fig.~\ref{f:filament}(c).
The value of $C_m$ with $m = 4$ is largest for $t \simeq 100$ in
Fig.~\ref{f:fourier}(b), which gives the modulation of the rings shown in 
Fig.~\ref{f:filament}(c).

We note that the significant growth of $C_4$ occurs roughly periodically,
e.g., at $t \simeq 40$ and $t \simeq 75$, as indicated by the vertical
dashed lines in Fig.~\ref{f:fourier}(b).
The vortex-ring configurations at $t = 40$ and $t = 75$ are shown as the
insets in Fig.~\ref{f:fourier}(b).
We find that the three vortex rings align in a rectilinear manner at these
instants, and the two vortex rings in the back are about to pass through the
front vortex ring.
Figure~\ref{f:fourier}(b) indicates that such dynamics significantly
increase the symmetry-breaking modulation of the vortex rings.
For the triangular initial arrangement, on the other hand, such a
rectilinear configuration of the three vortex rings is not realized in the
time evolution, and hence $C_m$ does not grow significantly.
The physical explanation for why such a vortex configuration is unstable has
yet to be clarified.

\begin{figure}[tb]
\includegraphics[width=8cm]{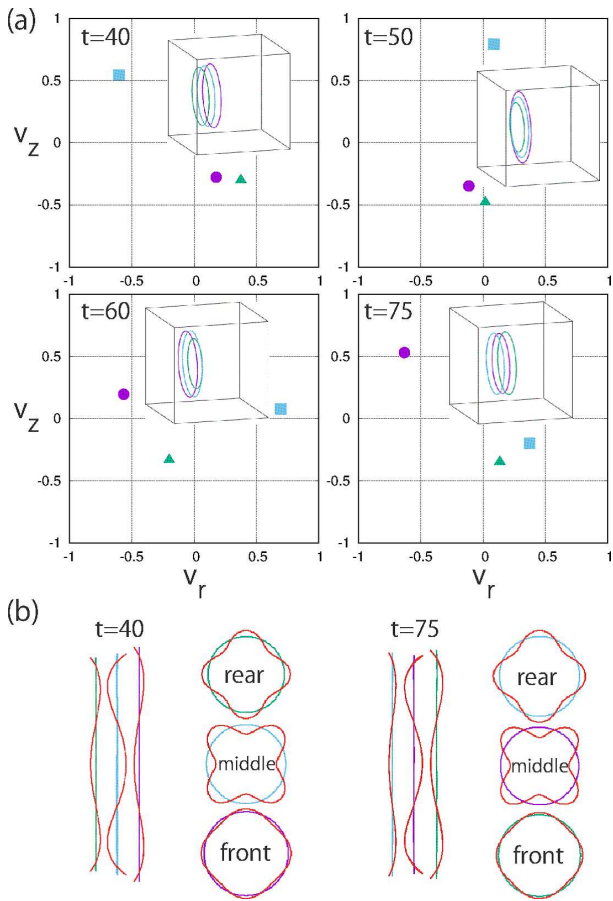}
\caption{
  (color online) The most unstable eigenvector of Eq.~(\ref{eigen}) for $m =
  4$.
  The three vortex rings are initially in the rectilinear configuration, as
  in Fig.~\ref{f:filament}(c).
  (a) The most unstable eigenvectors, $(c_4^{(1)}, d_4^{(1)})$, $(c_4^{(2)},
  d_4^{(2)})$, and $(c_4^{(3)}, d_4^{(3)})$ are plotted at each time, where
  the circles, triangles, and squares correspond to the vortex rings in the
  inset with the same colors (gray scales).
  (b) Shape of the most unstable mode with $m = 4$ at $t = 40$ and $t = 75$.
  The three vortex rings are seen from the $+x$ direction in the left-hand
  panel at each time.
  In the right-hand panels, the three vortex rings seen from the $+z$
  direction are separately shown for clarity.
}
\label{f:mode}
\end{figure}
To study the unstable modes, we perform a linear stability analysis.
The symmetry-breaking modulation of the $m$th mode is represented by the
Fourier coefficients in Eqs.~(\ref{cmn}) and (\ref{dmn}), which we write as
a vector $\bm{v}_m = (c_m^{(1)}, d_m^{(1)}, c_m^{(2)}, d_m^{(2)}, c_m^{(3)},
d_m^{(3)})^T$, with $T$ being the transpose.
When the modulation is small, we can linearize the equation of motion with
respect to $\bm{v}_m$, and therefore, we can linearize the time evolution of
the modulation as
\begin{equation} \label{eigen}
\bm{v}_m(t) = M(t) \bm{v}_m(0),
\end{equation}
where $M(t)$ is a $6 \times 6$ matrix.
The modes with different $m$ are not coupled with each
other~\cite{Widnall}.
We obtain the matrix $M(t)$ numerically as follows.
First, we set the initial vector as $\bm{v}_m(0) = (\epsilon, 0, 0, 0, 0,
0)^T$ and make the initial vortex rings using Eq.~(\ref{invF}), where
$\epsilon \ll 1$.
After the time evolution of the vortex rings, $\bm{v}_m(t)$ is obtained with
Eqs.~(\ref{cmn}) and (\ref{dmn}), which gives $(M_{11}(t), M_{21}(t), \cdots,
M_{61}(t))^T = \bm{v}_m(t) / \epsilon$.
In a similar manner, we obtain all the matrix elements of $M(t)$.

The eigenvector of $M(t)$ with the eigenvalue having the largest magnitude
corresponds to the most unstable mode.
Figure~\ref{f:mode} shows the form of the most unstable eigenvector for $m
= 4$ with the triangular initial arrangement as in Fig.~\ref{f:filament}(c).
All the elements of the eigenvector can be taken to be real.
In Fig.~\ref{f:mode}(a), we plot the most unstable mode on the $c_m$-$d_m$
plane.
At $t = 40$, the square (corresponding to the middle vortex ring) is located
opposite the circle and triangle (corresponding to the front and rear
vortex rings) across the origin.
These plots move in time as the leapfrog dynamics of the vortex rings, and
at $t = 75$, the circle (corresponding to the middle vortex ring) is located
opposite the triangle and square (corresponding to the front and rear
vortex rings) across the origin.
These eigenvectors of the unstable mode are visualized in
Fig.~\ref{f:mode}(b), where the modulation is superimposed on the
unmodulated vortex rings in an exaggerated manner.
We find that the modes at $t = 40$ and $t = 75$ have similar shapes, despite
the order of the vortex rings changing.
At both $t = 40$ and $t = 75$, the deviations from the unperturbed vortex
rings in adjacent vortex rings are opposite to each other.

\begin{figure}[tb]
\includegraphics[width=8cm]{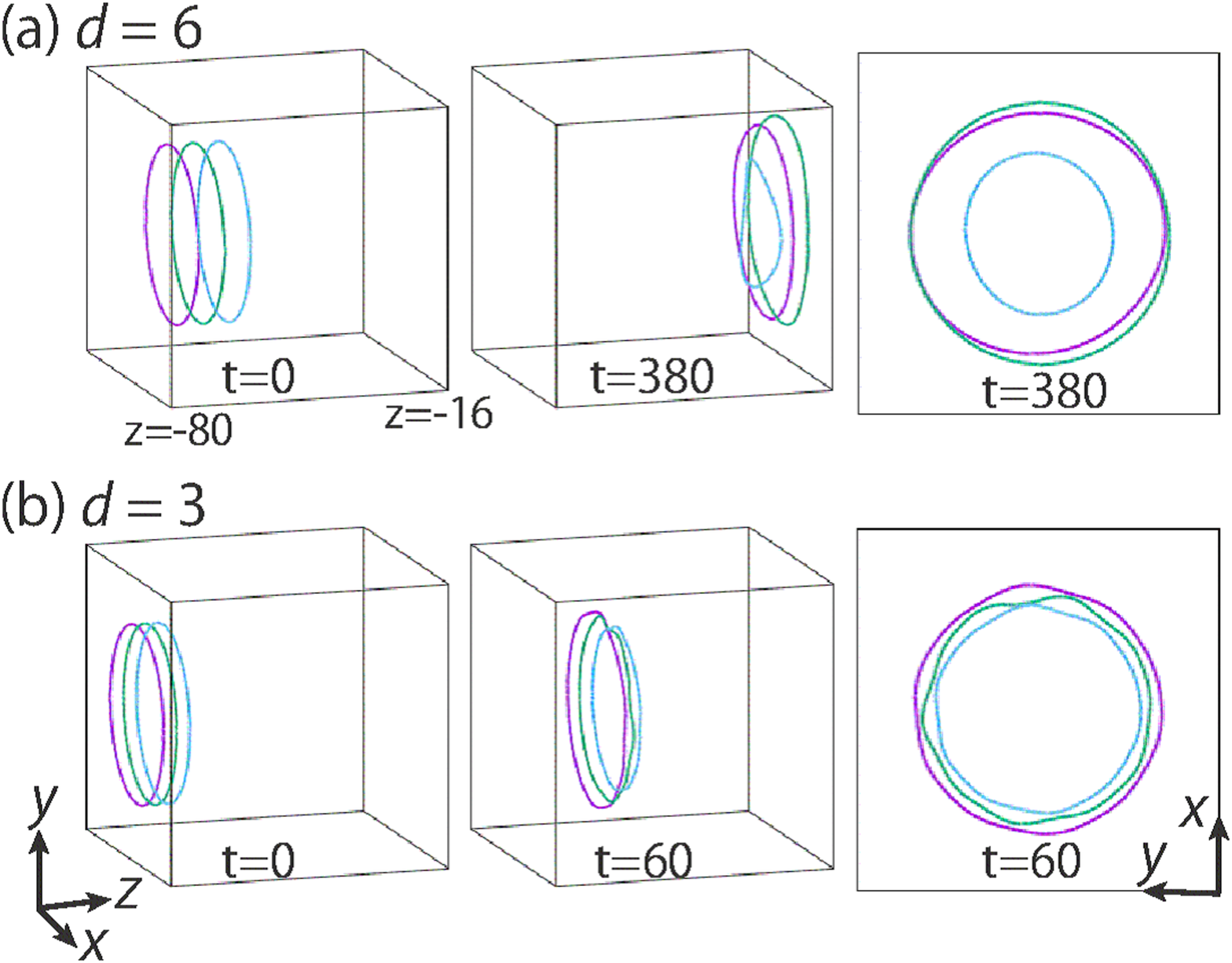}
\caption{
  (color online) Dependence of symmetry breaking dynamics on initial
  distance $d$ between vortex rings for rectilinear initial
  arrangement with radius $R = 20$.
  The dynamics are obtained by the vortex-filament model.
  (a) $d = 6$ with $Z_1 = -70$, $Z_2 = -64$, and $Z_2 = -58$.
  (b) $d = 3$ with $Z_1 = -30$, $Z_2 = -27$, and $Z_2 = -24$.
  The size of the box is $64^3$ with the $z$ axis at the center.
  $-80 < z < -16$ is shown at $t = 0$ in (a) and $-32 < z < 32$ is shown in
  the other boxes.
  The rightmost panels show the system seen from the $+z$ direction.
  See the Supplemental Material for movies showing the
  dynamics~\cite{movies}.
}
\label{f:d-dep2}
\end{figure}
Figure~\ref{f:d-dep2} shows the dependence of the axial symmetry breaking on
the initial distance $d$ between the vortex rings for the rectilinear
initial arrangement.
In Fig.~\ref{f:d-dep2}(a), the initial distance is $d = 6$, which is larger
than $d = 4$ in Fig.~\ref{f:filament}(c).
The axial symmetry breaking becomes significant at $t \simeq 300$ and the
modes $m = 2$ or 3 seem to be the most unstable.
For $d = 3$, as shown in Fig.~\ref{f:d-dep2}(b), the modulation with $m
\simeq 8$ becomes significant at $t \gtrsim 60$.
Thus, the wavelength of the symmetry-breaking modulation increases and the
growth rate of the modulation decreases with increasing $d$, which agrees
with the tendency in the GP model shown in Fig.~\ref{f:d-dep}.

\section{Conclusions}
\label{s:conc}

We have investigated the dynamics of coaxially arranged multiple vortex rings
using the GP model and vortex-filament model.
In both models, three vortex rings with rectilinear initial arrangement
(Fig.~\ref{f:gpevo}(d), right-hand panel) are found to be very unstable, and
the axial rotational symmetry of the system is spontaneously broken within a
few leapfrogs, as shown in Figs.~\ref{f:gpevo}(c) and \ref{f:filament}(c).
In contrast, three vortex rings with a triangular initial arrangement
(Fig.~\ref{f:gpevo}(d), left-hand panel) are much more stable, as shown in
Figs.~\ref{f:gpevo}(b) and \ref{f:filament}(b).
In the GP model, we have shown that the most unstable wavelength depends on
the initial distance between the vortex rings (Fig.~\ref{f:d-dep}) and the
interaction coefficient (Fig.~\ref{f:g-dep}).
In the vortex-filament model, we performed a Fourier analysis of the
modulation of the vortex rings, and found that the symmetry-breaking
modulation significantly grows when the three vortex rings are arranged in
a line and the rear vortex rings are about to pass through the front ring
(Fig.~\ref{f:fourier}(b)).
The shape of the unstable mode was obtained (Fig.~\ref{f:mode}).

Such symmetry-breaking dynamics of multiple vortex rings are difficult to
observe in experiments, since coaxial vortex rings with axial rotational
symmetry must be created in a controlled manner.
Phase imprinting~\cite{Ruo,Ruo2} and dynamical~\cite{Pinsker,Yukalov}
methods may realize such an arrangement of quantized vortex
rings in an atomic BEC.

\begin{acknowledgments}
We thank M. Tsubota and S. Yui for their valuable comments on the
vortex-filament method.
This work was supported by JSPS KAKENHI Grant Numbers JP17K05595,
JP17K05596, and JP16K05505.
\end{acknowledgments}

\end{document}